# Effect of the Surface States on Photoluminescence from Surface GaN/$Al_{0.2}Ga_{0.8}N$ Quantum Wells

Y. D. Glinka,[1,2*] T. V. Shahbazyan,[3] H. O. Everitt,[2,4] J. Roberts,[5] P. Rajagopal,[5] J. Cook,[5] E. Piner,[5] and K. Linthicum[5]

[1]*Institute of Physics, National Academy of Sciences of Ukraine, Kiev, 03028, Ukraine*
[2]*U.S. Army Aviation and Missile RDEC, Redstone Arsenal, Alabama 35898, USA*
[3]*Department of Physics, Jackson State University, Jackson, Mississippi 39217, USA*
[4]*Department of Physics, Duke University, Durham, North Carolina 27708, USA*
[5]*Nitronex Corporation, Raleigh, North Carolina 27703, USA*

We report on photoluminescence (PL) measurements at 85 K for GaN/$Al_{0.2}Ga_{0.8}N$ surface quantum wells (SQW's) with a width in the range of 1.51 – 2.9 nm. The PL spectra show a redshift with decreasing SQW width, in contrast to the blueshift normally observed for conventional GaN QW's of the same width. The effect is attributed to a strong coupling of SQW confined exciton states with surface acceptors. The PL hence originates from the recombination of surface-acceptor-bound ($A_s^0 X_A$) excitons. Two types of acceptors were identified.

It is well known that a decrease in the width of the semiconductor quantum wells (QW's) formed in the epitaxial stacks leads to an increase in the energy of quantum states which appears as a blueshift of the corresponding photoluminescence (PL) band.[1] In such quantum systems, the medium confining the well creates barriers, which prevent wave functions of electronic states from extending beyond the QW. If a QW is located at the surface of the epitaxial stack [surface QW (SQW)], the vacuum level at the surface of the QW material is assumed to play the role of such a barrier. However, the modeling of the surface as an abrupt termination of the structure with a quasi-infinite potential barrier has unambiguously been shown to be in a disagreement with experimental observations.[2] The study of ultra-high-vacuum PL from the GaAs/$Al_{0.3}Ga_{0.7}As$ QW's revealed a strong quantum coupling of the QW confined states to the surface states.[2] As a result, the PL from the GaAs QW's, confined by a thick $Al_{0.3}Ga_{0.7}As$ barrier on one side and capped by $Al_{0.3}Ga_{0.7}As$ barrier with thickness varied from 0 to 100 nm on the other side, showed a redshift and a decrease in intensity as the $Al_{0.3}Ga_{0.7}As$ cap layer width decreased, i.e., with surface states approaching to the QW states. The effect of relaxed surfaces subjected to air on the SQW states is expected to be more significant because of surface donor-like and acceptor-like centers associated with some impurities and structure defects.[3] However, the general tendency is expected to be similar to that for ultra-high-vacuum PL measurements since the difference is only in the nature of the surface states. Definitely, a comparison of the effect of the surface states on PL from GaAs and GaN SQW's is of a great importance since there are more polar bonds in GaN as compared to GaAs and hence the internal fields caused by spontaneous and piezoelectric polarizations should be taken into account for GaN SQW's.[4-6]

In the current paper, we report on PL from the GaN/$Al_{0.2}Ga_{0.8}N$ SQW's excited with the photon energy of 3.86 eV, which is just above the $Al_{0.2}Ga_{0.8}N$ barrier band-gap (3.8 eV). The PL from SQW's has been compared to that from a conventional $Al_{0.2}Ga_{0.8}N$/GaN/$Al_{0.2}Ga_{0.8}N$ QW measured under the same experimental conditions. With decreasing SQW width in the range of 2.9 – 1.51 nm, we observed a redshift of the PL bands of the order of 10 meV. Since in the conventional QW's, a decrease in the QW width of the same order leads to the blueshift,[4-6] the observed PL redshift points to a coupling of the SQW confined exciton state to the surface acceptor states. The PL hence originates from the recombination of surface-acceptor-bound excitons ($A_s^0 X_A$). Our findings are different from those obtained with more energetic excitations[7,8] (photon energy higher than 4 eV), where PL spectra are likely to reflect, in addition, the nonequilibrium carrier and phonon dynamics.

The GaN SQW's of different thicknesses (1.51, 1.6, 1.65, 1.7, 2.15, and 2.9 nm) were grown on a 100 nm Si(111) wafer in metalorganic chemical vapor deposition (MOCVD) reactor at nominally 1000 $^{o}$C.[7] The epitaxial stack consisted of a (Al,Ga)N-based transition layer, followed by ~ 800 nm of unintentionally doped (UID) GaN. The device layer consisted of ~ 31.5 nm of UID $Al_{0.2}Ga_{0.8}N$, which was capped with a thin UID GaN layer of different thicknesses mentioned above. The device layer for the conventional QW consisted of ~ 8 nm of UID $Al_{0.2}Ga_{0.8}N$ and 4 nm UID GaN, which was capped with 10 nm UID $Al_{0.2}Ga_{0.8}N$. All PL measurements were carried out in a vacuum temperature-controlled cryostat at 85 K. An optical parametric amplifier, pumped by a 1 kHz regenerative amplifier seeded by an 80 MHz Ti:Sa oscillator operating at 790 nm (170 fs pulses) in combination with the Topas light conversion system emitting the 321 nm (3.86 eV) light of an average power of 0.3 - 0.8 mW has been used as a

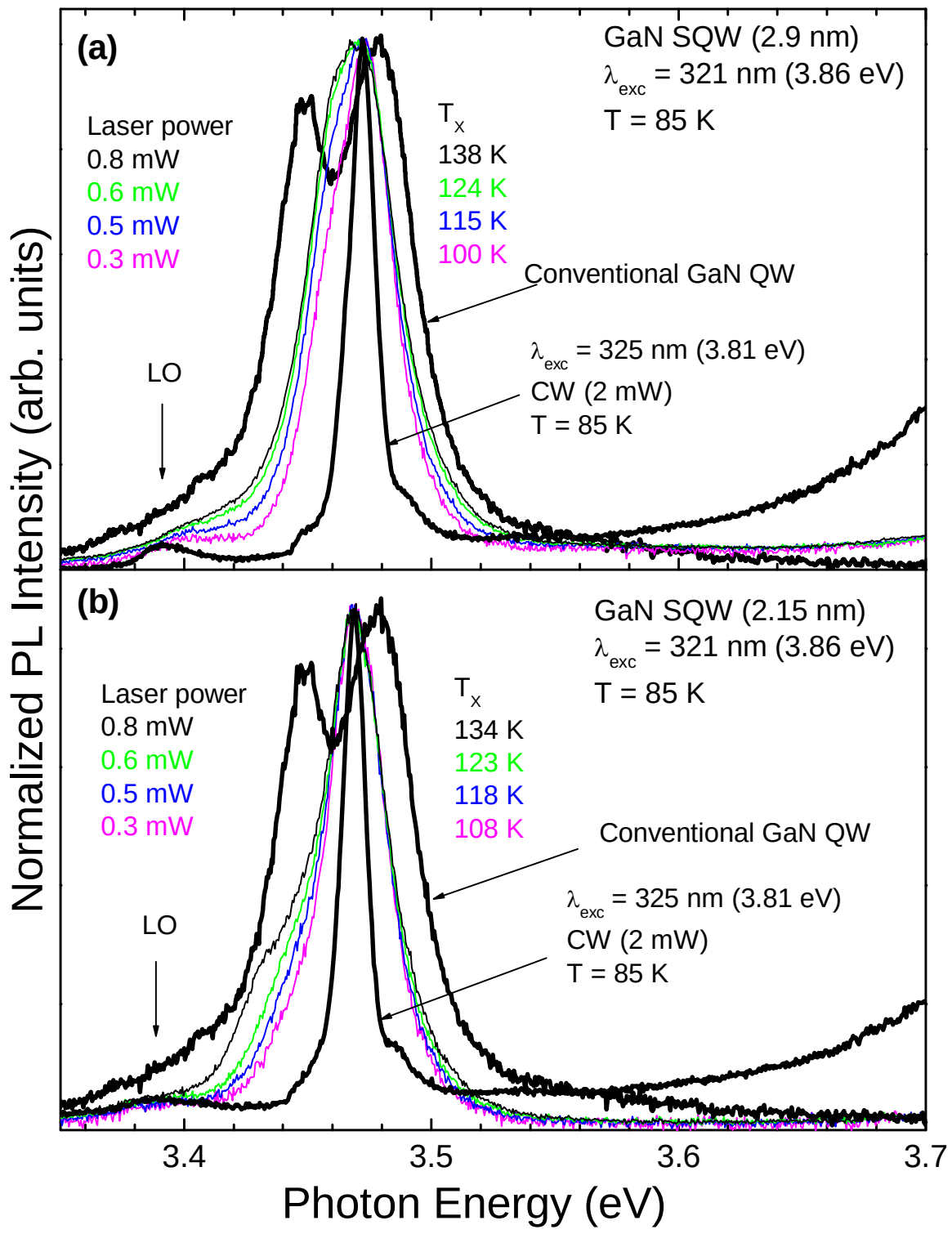

FIG. 1 (color online). PL spectra for GaN/$Al_{0.2}Ga_{0.8}N$ SQW's of widths 2.9 (a) and 2.15 nm (b) measured at 85 K with 3.86-eV photon energy and different average power (marked by different colors). The corresponding exciton temperatures ($T_x$) are also indicated. For comparison, the PL spectrum for conventional QW measured at 85 K with the same photon energy and 0.8-mW average power and the spectra for SQW's measured with unfocused CW 3.81-eV laser light (2 mW), are shown in both panels.

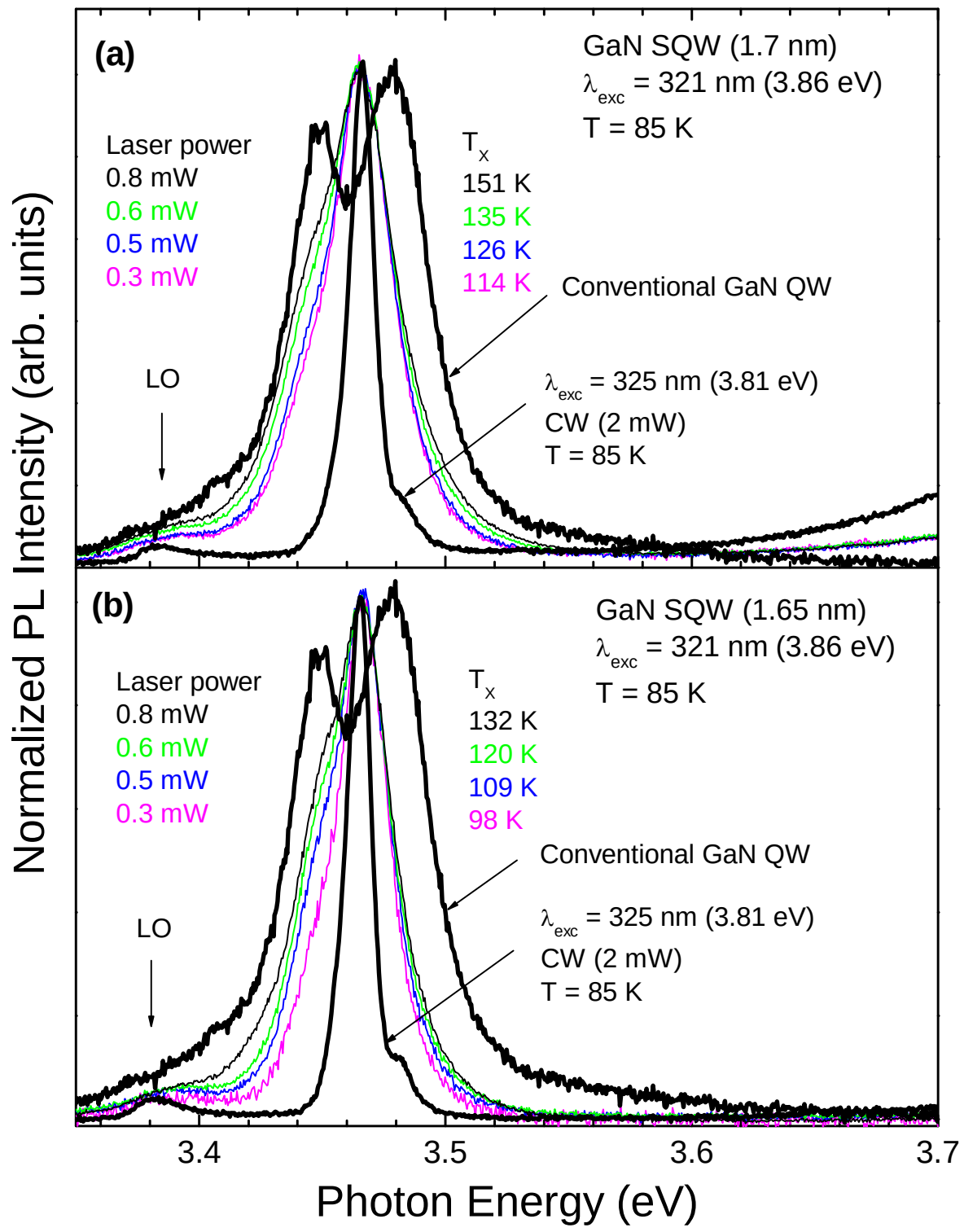

FIG. 2 (color online). Same as in Fig. 1 but for GaN/$Al_{0.2}Ga_{0.8}N$ SQW's of widths 1.7 (a) and 1.65 nm (b).

source for PL excitation. Also, the PL spectra measured with unfocused light from the continuous wave (CW) He-Cd laser (325 nm – 3.81 eV) of an average power of 2 mW has been used for comparison. The PL response was monitored either by a charge-coupled device camera or by a streak camera through the fiber optics and monochromators. The streak camera temporal resolution edge was 30 ps.

PL spectra for GaN SQW's of different widths and PL spectrum for the conventional GaN QW measured with pulse excitation are shown in Figs. 1 and 2 together with spectra for GaN SQW's measured with CW laser excitation. The spectrum for the conventional QW consists of two bands peaked at around 3.449 and 3.478 eV, which are assigned to the QW and bulk GaN (GaN buffer layer) excitonic emissions, respectively. The QW transition energy is lower than the bulk GaN excitonic emission energy due to the quantum-confined Stark effect caused by the strong built-in electric field in wurtzite GaN/AlGaN heterostructures.[4-6] In contrast, the PL from GaN SQW's exhibits features that cannot be explained by combination of the quantum confinement and quantum-confined Stark effects. The dominant PL band is peaked at energy (3.472 eV for 2.9 nm SQW) that is below the bulk GaN excitonic emission (the latter one is expected to be weak for these samples), and progressively redshifts (~ 10 meV) as the SQW width decreases (Fig. 3). The longitudinal optical (LO)-phonon sideband of the main peak and the less intense shoulder peak can be seen at lower and higher energies, respectively, when the CW laser excitation is applied (Figs. 1-3). The shoulder peak, which is distanced from the dominant PL peak by ~ 16 meV, shows a similar redshift (Fig. 3).

The energy positions of the PL peaks are different from those observed for conventional QW's, which are known to emit above the band-gap energy of bulk GaN if their width is smaller than 3.0 nm.[4-6] We attribute this difference to an additional effect of the surface states on the SQW confined exciton states. Here we suppose that the built-in electric field plays a significant effect on GaN SQW's. Because the piezoelectric field is known to mainly affect the AlGaN barriers and not the GaN layer,[5] its effect on SQW's is weak. However, this is not the case for the spontaneous polarization field, which is expected to give a dominant contribution into the quantum-confined Stark effect in SQW's.[6] Nevertheless, we show below that the redshift of the PL band due to the quantum-confined Stark effect cannot explain the PL redshift observed. This is also supported by the fact that the conventional QW's thinner than 3.0 nm show a blueshift of excitonic PL, i.e., the quantum confinement effect for these QW widths dominates over the quantum-confined Stark effect.[4-6] The space-charge field originating from the Fermi-level pinning

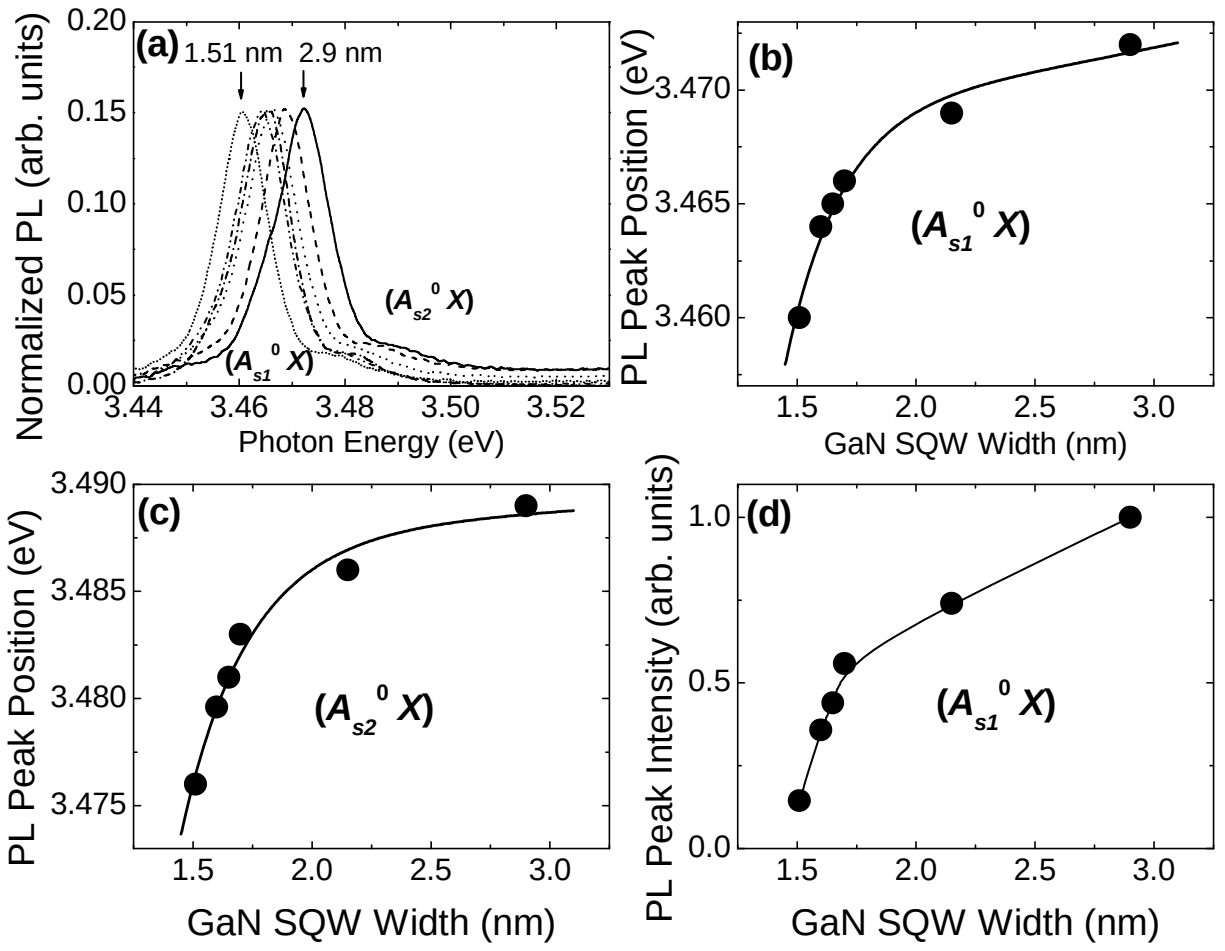

FIG. 3. PL spectra measured at 85 K with CW 3.81-eV laser excitation (2 mW) are shown for different GaN/$Al_{0.2}Ga_{0.8}N$ SQW's widths (a), together with the corresponding width dependences of the ($A_{s1}^0X_A$) exciton peak position (b), the ($A_{s2}^0X_A$) exciton peak position (c), and the ($A_{s1}^0X_A$) exciton peak intensity (d). The curves in (b) and (c) show the fit to the data in assumption that the free exciton band is peaked at 3.5 eV. The curve in (d) is a guide to the eye.

at the surface can be disregard since it changes on a much larger length scale of ~ 100 nm. Thus we can safely neglect all field effects with exception of the spontaneous polarization field and argue that the GaN SQW confined exciton state should be coupled to surface states, in a similar way as in GaAs SQW's.[2] Specifically, the dominant PL band and the shoulder peak originate from the recombination of surface-acceptor-bound excitons, which are bound to two types of surface acceptors: ($A_{s1}^0X_A$) and ($A_{s2}^0X_A$), respectively. At the temperature used (85 K), the effect of surface neutral donors is negligible due to the much smaller binding energy of excitons bound to donors.[3] Here we stress that the main distinctive feature of GaN SQW's, as compared to the conventional QW's, is a high density of surface acceptors (~ $10^{12}$ $cm^{-2}$).[9] One can estimate the sheet exciton density in SQW's by taking into account the Beer's law distribution of carriers photoexcited in $z$ direction (perpendicular to the SQW plane) at the maximal laser intensity applied ($9.4\times10^{10}$ W/$cm^2$) and the GaN material parameters. The power density absorbed in the media is $P = -\operatorname{div} S$, with $S = I_0(1-R)\exp(-\alpha z)$, where $I_0$ is the initial laser intensity at the sample surface, $R$ = 0.2 (for photon energy of 3.86 eV) is the normal incidence reflectivity, and $\alpha$ = $1.0\times10^5$ $cm^{-1}$ is the absorption coefficient for photon energy of 3.86 eV. Hence, the power density absorbed within the absorption length is $P = \alpha I_0(1-R)\exp(-\alpha z)$ = $2.76\times10^{13}$ W $cm^{-3}$. The resulting density of excitons photoexcited in the media of the absorption length width is $7.6\times10^{18}$ $cm^{-3}$, which corresponds to the $8.3\times10^{12}$ $cm^{-2}$ sheet exciton density photoexcited in the SQW. The latter value is comparable to the surface acceptor density and so the surface-acceptor-bound excitons are expected to dominate the light emission process. This is

a

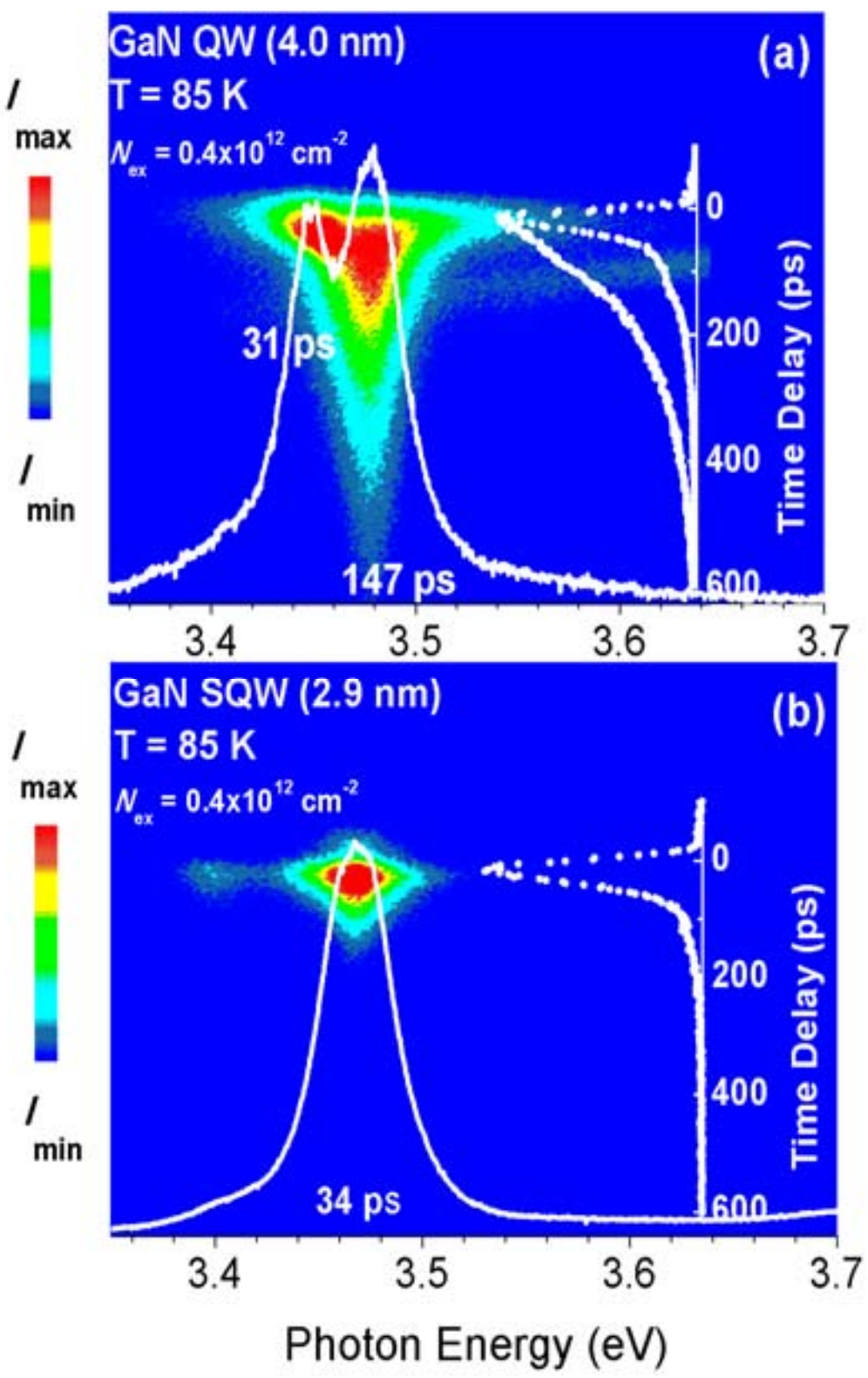

FIG. 4. (color online). Examples of streak-camera PL patterns for GaN/$Al_{0.2}Ga_{0.8}N$ conventional QW (a) and SQW (b) samples. PL spectra correspond to those shown in Fig. 1.

a reason why the PL from ($A_s^0X_A$) excitons dominates the emission spectrum at 85 K. We do not discuss here the identity of these acceptors by noting only that the energy difference between ($A_{s1}^0X_A$) and ($A_{s2}^0X_A$) peaks well matches the range of differences for typical binding energies of excitons bound to deep acceptors in GaN.[3] As this takes place, the PL from other sources is expected to be much weaker. For instance, the contribution to the PL from 2D electron gas, which can possibly be formed at the GaN buffer layer/$Al_{0.2}Ga_{0.8}N$ barrier interface, is expected to be relatively small, because of a large spatial separation of 2D electrons and holes localized at bulk acceptors.[10]

The assignment of the observed PL to ($A_s^0X_A$) excitons in SQW's is also supported by time-resolved measurements (Fig. 4 and 5). In order to control the streak camera temporal resolution edge (25 ps), we also measured PL spectra with higher laser power, which revealed a stimulated emission feature appeared in the lower energy range additionally to the PL bands mentioned above.[11] As follows from Figures 4 and 5, the PL decay for conventional QW and SQW's is close to the streak camera temporal resolution edge. As an example, we point that the decay time measured for 2.9 nm SQW is 34 ps while for thinner SQWs this time becomes even shorter and so follows the temporal resolution edge. This is consistent with the time-resolved measurements of the GaN multiple-quantum-well structures.[11] In contrast, the PL from the bulk GaN (buffer layer) in the conventional QW sample shows much longer decay of 147 ps. The difference is attributed to the higher rate of nonradiative recombination in QWs and SQWs as compared to the bulk GaN, which we associate with impurities and structure defects at the interfaces. We stress here that the time-resolved measurements allow us to distinguish between the buffer layer PL from the conventional QW sample and the SQW ($A_s^0 X_A$) exciton PL, despite the small difference in their peak positions (Figs. 1, 2, 4, and 5).

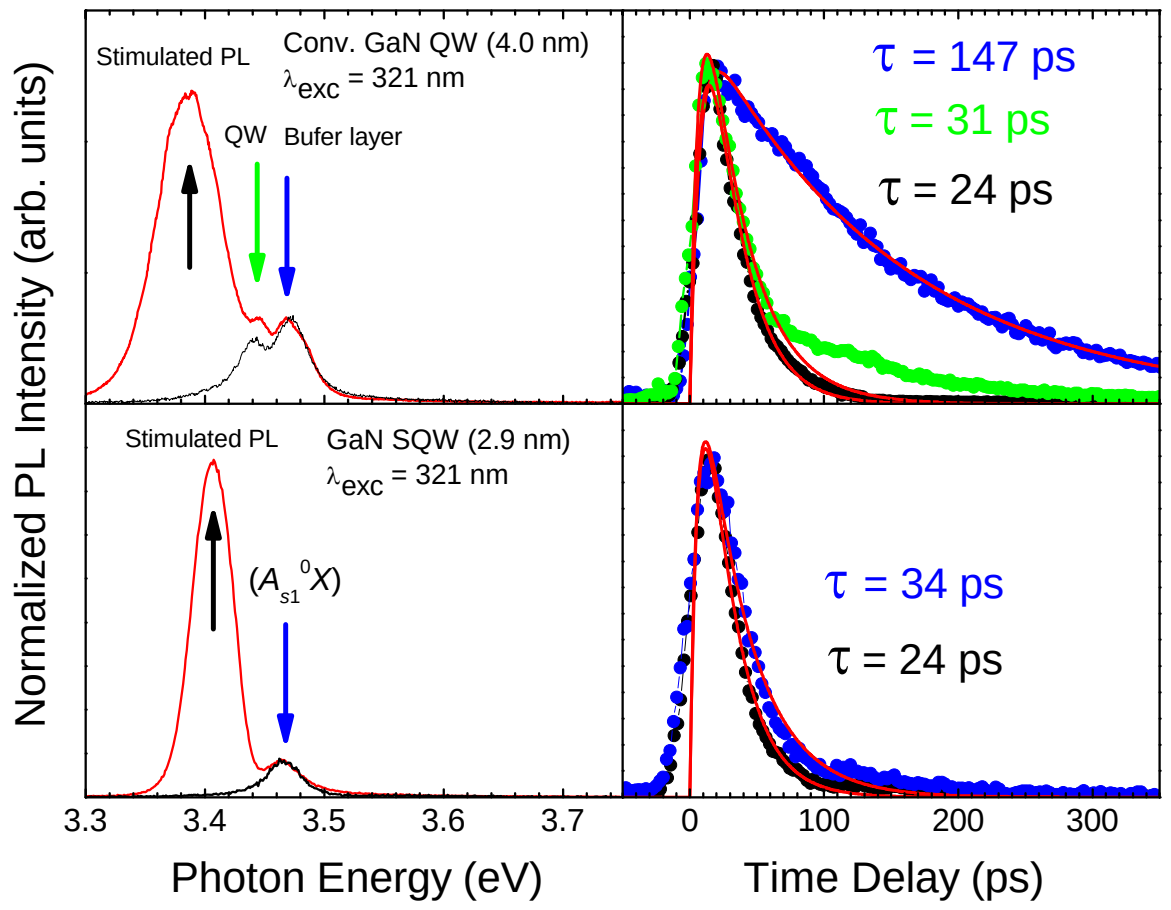

FIG. 5. (color online). (left column) PL spectra for conventional QW of 4 nm width and SQW of 2.9 nm width measured at 85 K with the average laser power of 0.8 mW (black curves) and 8 mW (red curves). (right column) The corresponding PL decays measured for PL peaks indicated in the left column by the corresponding color arrows. The red thin curves show the fit to the data.

Thus the observed redshift of ($A_s^0 X_A$) exciton PL is caused by the quantum coupling between a neutral acceptor at the surface and the exciton state in the SQW. The actual shift of PL peak is hence determined by a competition of several confinement-induced effects. The first effect is the upward shift of electronic levels with decreasing GaN SQW width due to the quantum confinement effect: $E_{PL} = E_g + \left(1/m_e + 1/m_{hh}\right)\hbar^2\pi^2/2L_W^2$, where $E_g$ is the GaN bandgap energy, $m_e = 0.2\ m_0$ and $m_{hh} = 1.64\ m_0$ are the in-plane electron and heavy hole effective masses ($m_0$ is the free electron mass), and $L_W$ is the SQW width.[4-6,12] On the other hand, the quantum-confined Stark effect, which is caused by the strong built-in electric field originating from spontaneous polarization in SQW's, gives the following PL energy: $E_{PL} = E_g - pF + \beta F^2$, where $p$ is a permanent dipole moment of an exciton, $F$ is the spontaneous polarization field $F = \Delta V_0/L_W$ ($\Delta V_0$ denotes the difference of potentials at the surface and at the $GaN/Al_{0.2}Ga_{0.8}N$ interface), and $\beta$ is a polarizability of electrons and holes[13] $\beta = e^2 m_{e,hh} L_{e,hh,z}^4/2\hbar^2$ with $L_{e,hh,z} = \sqrt{\hbar/m_{e,hh}\omega_{e,hh,z}}$ being an extent of electron and hole ground state wave functions in the direction perpendicular to the SQW plane, $z$ (the indexes are applied for either the electron ($e$) or the heavy hole ($hh$), respectively, $\omega_e$ and $\omega_h$ are the corresponding frequencies, and $e$ is the electron charge). In other words, the different confining potentials for electron and holes lead to their spatial separation creating the permanent dipole momentum which interacts with the field. This effect gives a redshift of the excitonic PL band, which is linear with $F$ and so scales as $1/L_W$. Also, the built-in electric field pulls the electron and hole wave functions apart inducing the corresponding blueshift, which scales as $1/L_W^2$. Finally, one should take into account a decrease of the exciton Bohr radius ($a_B$ = 2.7 - 2.8 nm,[12,14]) with decreasing $L_W$ in the range of 2.9 - 1.51 nm, which, consequently, leads to an increase in the exciton binding energy and causes the redshift of the excitonic PL, which scales as $1/L_W$. As this takes place, the effect of the screening of the polarization field by free carriers on the exciton binding energy is assumed to be weak due to the small density of the free carriers photoexcited with CW excitation in the SQWs. For free excitons, the blueshift is known to dominate over the redshift, resulting in an overall blueshift of the excitonic PL band.[12] The reason is that the Stark, Coulomb, and dipole energies scale only as $1/L_W$, while the quantum confinement and polarization effects scale as $1/L_W^2$. However, the situation is even more complicated in QW's heavily doped with donor or acceptor impurities, where PL is dominated by bound exciton complexes. Systematic studies of neutral-donor-bound exciton PL revealed a non-monotonic dependence of PL peak position on the QW width.[15] With a reduction of the QW width, the binding energy of the exciton bound to a neutral donor exhibits a maximum at $L \sim a_B$ and then decreases.[15] For narrow QW's, this effect prevails over confinement-induced upward shift of electron levels and results in the overall redshift of PL band.

We observe a similar behavior in SQW's as well due to the high concentration of surface acceptors. In this case, the excitons are bound to a sheet of acceptors that are located in a close proximity to the GaN SQW. Since the interaction between free exciton and neutral acceptor is short-ranged, the binding energy of the ($A_s^0 X_A$) exciton is very sensitive to the ratio between the effective Bohr radius of a neutral acceptor and the SQW width. The effective Bohr radius of an acceptor $a_A = \hbar/\sqrt{2m_{hh}E_A}$, where $E_A$ is the acceptor ionization energy,[16] can be estimated for most common neutral acceptors in UID GaN [Ref. 3] as 0.40 nm (C), 0.34 nm (Si), 0.31 (Mg), and 0.24 nm (Zn). Because the built-in electric field aligns excitons along the $z$ direction, the shortening of the interaction scale between excitons and surface neutral acceptors cannot be remedied by a reorientation of the exciton dipole moment with respect to the QW plane. This explains the sharp width dependence of PL redshift and intensity for narrow SQW's. The PL intensity drop with decreasing SQW width is caused by an increase in the nonradiative surface recombination.[17] This is consistent with the aforementioned shortening of the PL decay time with decreasing SQW width. The SQW width dependences of PL shift shown in Fig. 3 can be well fitted by adding to the aforementioned effects the binding energy effect and assuming that the binding energy of the ($A_s^0 X_A$) exciton has a dipole-dipole nature and hence scales as $1/L_W^3$. Note that the dependences of the PL peak position and PL intensity on the GaN SQW width (Fig. 3) are similar enough to those observed for GaAs QW's when $Al_{0.3}Ga_{0.7}As$ cap barrier thickness was varied from 100 to 0 nm.[2] This is consistent with the model discussed in the current paper.

The PL spectra measured with pulse excitation reveal a very large broadening (Fig. 1 and 2), which increases with laser power in the range of 0.3 – 0.8 mW (the photoexcited carrier density of $2.8\times10^{18}$ – $7.6\times10^{18}$ $cm^{-3}$). The peak position remains almost unchanged for different powers applied, but it redshifts with decreasing SQW width, similarly to PL spectra measured with CW laser excitation. The higher energy wing of the broadening shows an exponential behavior due to the hot exciton effect.[1,18,19] Assuming the PL spectral shape of the form[18] $I_{PL}(\hbar\omega)\propto(\hbar\omega-E_x)^{1/2}\exp\left[-(\hbar\omega-E_x)/(k_B T_x)\right]$, where $\hbar\omega$ is the emitted photon energy, $E_x$ and $T_x$ are the exciton energy and temperature, $k_B$ is the Boltzmann constant, the surface-acceptor-bound exciton temperature can be estimated by fitting the PL spectra measured (Figs. 1 and 2). The maximal temperature obtained is 151 K (~ 13 meV), i.e., less than the binding energy of excitons bound to the neutral acceptors and the exciton binding energy (26.3 meV).[3] The lower energy wing of the broadening results from the band-gap renormalization by the hot plasma, which progresses with the power density of excitation.[20]

In summary, we have provided evidence that PL from $GaN/Al_{0.2}Ga_{0.8}N$ SQW's results from the recombination of excitons bound to two types of surface acceptors. The red shift of PL with decreasing SQW width is explained as a result of the confinement-induced decrease in the binding energy of excitons bound to the surface acceptors in a close proximity to the GaN SQW.

One of the authors (Y.D.G.) gratefully acknowledges support from the U. S. Army Aviation and Missile RDEC. We thank Dr. Muth for helpful comments on this work. The work at Jackson State University was supported in part by the NSF under Grants DMR-0906945 and HRD-0833178.

* E-mail: glinka@icnanotox.org